\documentclass[lettersize,journal]{IEEEtran}
\IEEEoverridecommandlockouts
\usepackage{cite}
\usepackage{amsmath,amssymb,amsfonts}
\usepackage{algorithmic}
\usepackage{graphicx}
\usepackage{subfig} 
\usepackage{caption}
\usepackage{textcomp}
\usepackage{xcolor}
\usepackage{stfloats}
\usepackage{cuted}
\usepackage{balance}
\def\BibTeX{{\rm B\kern-.05em{\sc i\kern-.025em b}\kern-.08em
    T\kern-.1667em\lower.7ex\hbox{E}\kern-.125emX}}

\begin{document}

\title{Communication-Oriented Channel Characteristics of Holographic Curvature-Reconfigurable Apertures\\
}

\author{
    Hang Lin, Shu Sun, \textit{Senior Member, IEEE}, Hangsong Yan, \textit{Member, IEEE}, Qiuming Zhu, \textit{Senior Member, IEEE}
\thanks{H. Lin and S. Sun are with the School of Information Science and Electronic Engineering, Shanghai Jiao Tong University, Shanghai 200240, China (e-mail: \{linhang7346, shusun\}@sjtu.edu.cn). 

H. Yan is with Hangzhou Institute of Technology, Xidian University, Hangzhou 311231, China (e-mail: yanhangsong@xidian.edu.cn).

Q. Zhu is with the College of Electronic and Information Engineering, Nanjing University of Aeronautics and Astronautics, Nanjing 211106, China (email: zhuqiuming@nuaa.edu.cn).
    
%(\textit{Corresponding author: Shu Sun.})
}
}

\maketitle

\begin{abstract}
This letter investigates the joint impact of array curvature and mutual coupling (MC) on holographic curvature-reconfigurable apertures.
An analytical framework is developed to characterize spatial correlation, coupling-aware reference directivity, and spectral efficiency.
Under the small-curvature regime, we establish a relation linking inter-element distances to coupling perturbations and spatial-mode eigenvalues.
Simulation results reveal asymmetric spectral-efficiency variations between transmit- and receive-side array curvature with MC, while their no-MC difference is consistent with zero under hemispherical isotropic scattering.
Results based on the 3rd Generation Partnership Project clustered delay line-B channel angular/power profile, together with colored-noise sensitivity analysis, further show that the effects of curvature and receive-side coupling depend on the propagation angular distribution and receiver-noise covariance.
These findings highlight the importance of jointly accounting for array geometry, mutual coupling, and receiver noise when evaluating curved holographic multiple-input multiple-output links.
\end{abstract}


\section{Introduction}
Holographic multiple-input multiple-output (HMIMO) uses densely sampled electromagnetic apertures to provide fine spatial resolution and flexible wavefront control \cite{10130641,11247918,10232975}.
Operation near the continuous-aperture regime changes the array response and spatial correlation relative to conventional MIMO \cite{10158997, 11513410}.
Related developments include spherical stacked intelligent metasurfaces (SIMs) for full-space wave-domain processing \cite{11474787} and near-field beam training for SIM-assisted communication \cite{11534974}.
Dense element placement also introduces mutual coupling (MC), which modifies the effective array response and cannot generally be neglected \cite{7776022,11006094}.
Meanwhile, conformal and surface-mounted deployments can require curved apertures \cite{10824216,11174411}.
Curvature changes both propagation phases and inter-element impedances; consequently, its performance impact cannot be inferred from geometry or fixed-array MC alone.

Motivated by the need to jointly characterize these geometry- and coupling-dependent effects from a communication perspective, this letter studies holographic curvature-reconfigurable apertures (HoloCuRAs) using an impedance-based framework for spatial correlation, coupling-aware reference directivity, and spectral efficiency.
We establish a non-additive curvature–MC interaction by deriving, under isotropic scattering, a second-order decomposition that separates curvature-induced correlation changes filtered by the existing coupling network from those caused by curvature-dependent updates of the coupling matrices. 
Using 1000 paired channel realizations, a coupled-minus-uncoupled bending contrast and a fixed-planar-MC control show that the interaction is governed primarily by how the existing coupling network changes the link sensitivity to curvature, while curvature-induced changes in the coupling matrices provide only a secondary contribution. 
Independent transmit (TX) and receive (RX) bending further reveals statistically distinct one-sided responses with MC, whereas the corresponding no-MC difference is consistent with zero; a 3rd Generation Partnership Project (3GPP) clustered delay line (CDL)-B angular/power profile and coupling-dependent noise results show how the magnitude and sign of this interaction depend on the propagation and receiver assumptions.
%We relate small curvature to second-order coupling and correlation-eigenvalue perturbations under isotropic scattering.
%Using independent transmit (TX) and receive (RX) curvatures and 1000 paired realizations, we find unequal one-sided bending responses with MC, while the no-MC difference is consistent with zero under isotropic scattering, white noise, and the adopted input normalization.
%A 3rd Generation Partnership Project (3GPP) clustered delay line (CDL)-B angular/power profile and coupling-dependent noise sensitivity further quantify the dependence of the observed trends on propagation and receiver assumptions.

\section{System Model}
\begin{figure}[b]
	\centering
	\includegraphics[scale=0.165]{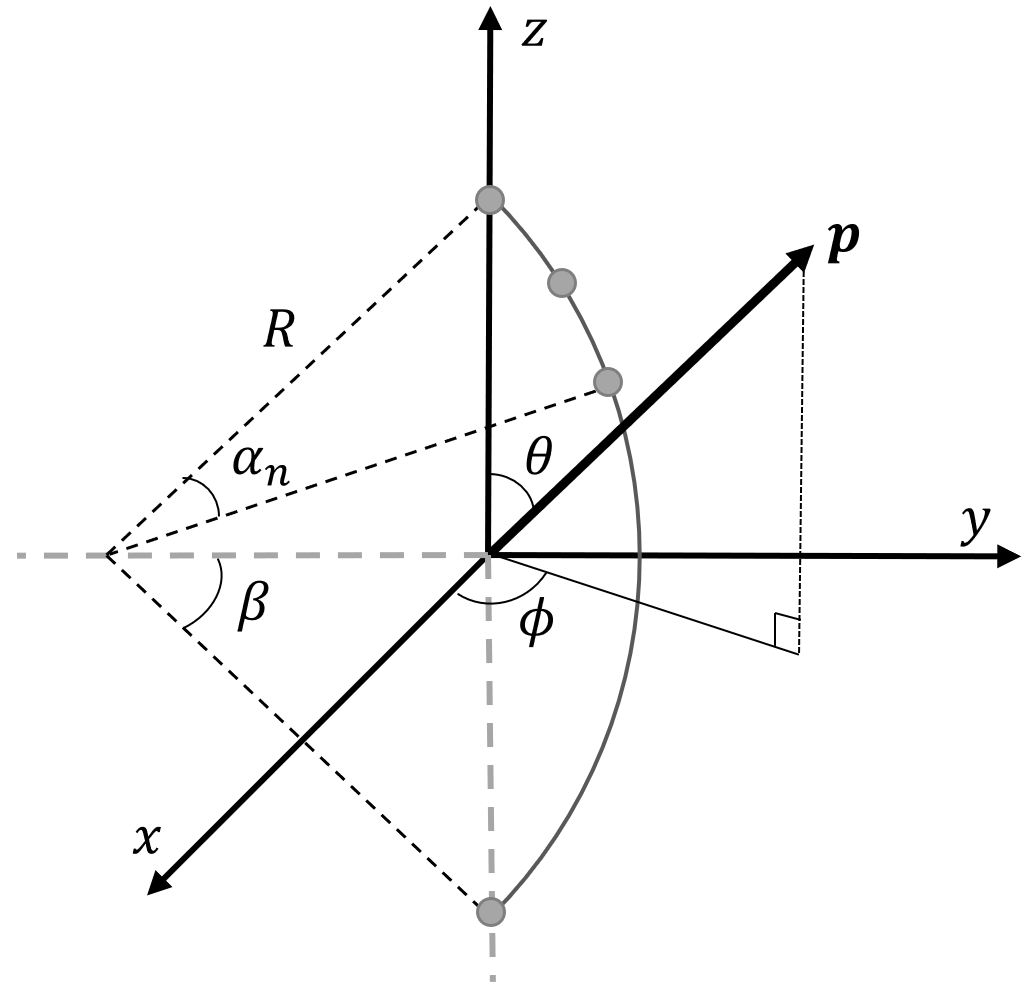}
	\captionsetup{font=footnotesize}
    \caption{1D HoloCuRA in the $yz$ plane.}  % 主图标题
	\label{FigFlexibleULAwithCoordinateSystem} % 主图标签
\end{figure}

We consider a one-dimensional (1D) HoloCuRA consisting of $N$ antennas, as depicted in Fig.~\ref{FigFlexibleULAwithCoordinateSystem}, where the array is bendable to form an arch-like shape.
The array has arc length $L$ and half opening angle $\beta \in [0,\frac{\pi}{2}]$, so its total opening angle is $2\beta$.
When $\beta = 0$, the array forms a straight line, corresponding to the special case of a uniform linear array (ULA).
%When $\beta = \frac{\pi}{2}$, the arc is exactly a semicircle.
The curvature radius of the array is $R = \frac{L}{2\beta}$.
The relative position of the $n$-th antenna element is described by the central angle $\alpha_n = \frac{(n-1)L}{(N-1)R},n=1,2,\ldots,N$.
The Cartesian coordinates of the $n$-th antenna element are written as
\begin{equation}
     \boldsymbol{r}_n = (0, R \cos{(\beta - \alpha_n)} - R \cos{\beta}, R \sin{(\beta - \alpha_n)}).
\end{equation}
The distance between the $n$-th antenna element and the $m$-th antenna element is calculated as
\begin{equation}
     {||\boldsymbol{r}_n - \boldsymbol{r}_m ||}_2 = 2R \sin{\frac{|\alpha_n - \alpha_m|}{2}}=\frac{L}{\beta} \sin{\frac{|n-m|}{N-1}\beta}.\label{spacing}
\end{equation}
For \(\beta\to0\), (\ref{spacing}) reduces to the ULA spacing \(\|\mathbf r_n-\mathbf r_m\|_2=|n-m|L/(N-1)\).
Letting $d_{nm}=\|\boldsymbol r_n-\boldsymbol r_m\|_2$ and $q_{nm}=|n-m|/(N-1)$, the small-curvature expansion of \eqref{spacing} is
\begin{equation}
 d_{nm}(\beta)=Lq_{nm}-\frac{Lq_{nm}^{3}}{6}\beta^2+\mathcal O(\beta^4).
 \label{small_curvature_spacing}
\end{equation}
Thus, for a mutual-impedance model that is smooth in $d_{nm}$, the resulting TX and RX coupling matrices satisfy
$\mathbf C_{T/R}(\beta)=\mathbf C_{T/R}(0)+\mathcal O(\beta^2)$.
This local relation applies when the network inverses remain nonsingular; it predicts a second-order MC perturbation at fixed aperture length.

The array is excited by the complex current vector $\boldsymbol{j}_0={(j_1,j_2,\ldots,j_N)}^{\mathrm{T}} \in \mathbb{C}^{N}$.
Considering the far-field situation \cite{10819602}, in Fig.~\ref{FigFlexibleULAwithCoordinateSystem}, $\boldsymbol{p}$ denotes the propagation vector, $\theta \in [0, \pi]$ is the zenith angle, and $\phi \in [0, 2\pi]$ represents the azimuth angle.
Given the current vector, the output signal of the array in the propagation direction is expressed as
\begin{equation}
    J_0(\phi, \theta) = \sum_{n = 1}^N j_n e^{j k \hat{\boldsymbol{p}} \cdot \boldsymbol{r}_n} = \boldsymbol{a}_0^{\mathrm{T}} \boldsymbol{j}_0, \label{OutputSignal}
\end{equation}
where $\hat{\boldsymbol{p}}=\left( \sin{\theta} \cos{\phi}, \sin{\theta} \sin{\phi}, \cos{\theta}\right)$ is the unit vector of the propagation path, $k$ denotes the wavenumber and $\boldsymbol{a}_0$ is the array steering vector
\begin{equation}
    \boldsymbol{a}_0 = {\left( e^{j k \hat{\boldsymbol{p}} \cdot \boldsymbol{r}_1},e^{j k \hat{\boldsymbol{p}} \cdot \boldsymbol{r}_2},\ldots,e^{j k \hat{\boldsymbol{p}} \cdot \boldsymbol{r}_N}\right)}^{\mathrm{T}}.
\end{equation}
%Preliminary expansion gives
%\begin{align}
%\begin{split}
%    \hat{\boldsymbol{p}} \cdot \boldsymbol{r}_n =& \sin{\theta} \sin{\phi} \cos{(\beta - \alpha_n)} -  \sin{\theta} \sin{\phi}\cos{\beta} \\
%   &+ \cos{\theta}\sin{(\beta - \alpha_n)}.
%\end{split}
%\end{align}

\subsection{Ignoring MC}
The spatial correlation matrix of the antenna array is\cite{7024192}
\begin{equation}
    \mathbf{R}_0=\int_0^\pi \int_0^{2\pi} s(\phi, \theta) \mathbf{a}_0(\phi, \theta) \mathbf{a}_0^{\mathrm{H}}(\phi, \theta) \mathrm{d} \phi \mathrm{d} \theta ,
\end{equation}
where $s(\phi, \theta)$ represents the normalized spatial scattering function.
For the isotropic scattering environment, we have
\begin{equation}
    s(\phi, \theta)=\frac{\sin{\theta}}{4 \pi}, \phi \in[0, 2\pi], \theta \in[0, \pi] .
\end{equation}

To obtain the correlation matrix in closed form, we use the integral identity
%(proved in the Appendix)
\begin{align}
\begin{split}
    &\int_{0}^{\pi} \int_{0}^{2\pi} e^{j \pi u \sin{\theta} \sin{\phi}} e^{j \pi v \cos{\theta}} \sin{\theta} \mathrm{d} \phi \mathrm{d} \theta \\
    &=4 \pi \operatorname{sinc}\left(\sqrt{u^{2}+v^{2}}\right),
\end{split}
\end{align}
where $\operatorname{sinc}(x)=\frac{\sin{(\pi x)}}{\pi x}$.
%Applying the above integral formula, 
The expression for the spatial correlation matrix can be obtained as
\begin{equation}
    \left[\mathbf{R}_0\right]_{n, m}=\operatorname{sinc}\left(\frac{2\left\|\boldsymbol{r}_n - \boldsymbol{r}_m\right\|_2}{\lambda}\right), n, m=1, \ldots, N , \label{SpatialCorrelationMatrix}
\end{equation}
where $\lambda$ is the carrier wavelength satisfying $\lambda = \frac{2\pi}{k}$.
The dominant eigenvalues of $\mathbf{R}_0$ indicate the usable spatial modes and the corresponding mode strengths.
%In MIMO channels, the spatial correlation matrix reveals usable spatial modes: the number of dominant nonzero eigenvalues approximates the channel’s spatial DoF and thus the number of parallel spatial streams.
%The eigenvalue magnitudes reflect the mode quality.
%consistent with the spectral-efficiency expression obtained from the nonzero singular modes of the MIMO channel.

The directivity in a specific direction $(\phi_0, \theta_0)$ is defined as the ratio of the power spectral density in that direction to its average over the full-space \cite{9838533}
\begin{equation}
    D_0\left(\boldsymbol{j}_0, \phi_0, \theta_0 \right)=\frac{4 \pi\left|J_0\left(\phi_0, \theta_0 \right)\right|^2}{\int_0^\pi \int_0^{2\pi}\left|J_0\left(\phi, \theta \right)\right|^2 \sin \theta \mathrm{d} \phi \mathrm{d} \theta}, \label{DefDirecitivty}
\end{equation}
where the factor $\sin{\theta}$ comes from the spatial integral form in spherical coordinates, and $4\pi$ is the normalization coefficient.

Note that \eqref{OutputSignal} yields $\left|J_0\left(\phi, \theta \right)\right|^2 = \boldsymbol{j}_0^{\mathrm{H}} \boldsymbol{a}_0^* \boldsymbol{a}_0^{\mathrm{T}} \boldsymbol{j}_0$.
It can be observed that the integral form in the denominator of \eqref{DefDirecitivty} is consistent with the spatial correlation matrix, from which a concise expression can be obtained
\begin{equation}
    D_0\left(\boldsymbol{j}_0, \phi, \theta \right)=\frac{\boldsymbol{j}_0^{\mathrm{H}} \boldsymbol{a}_0^* \boldsymbol{a}_0^{\mathrm{T}} \boldsymbol{j}_0}{\boldsymbol{j}_0^{\mathrm{H}} \mathbf{R}_0 \boldsymbol{j}_0}.
\end{equation}

For a given HoloCuRA, its maximum directivity in a specific direction $(\phi, \theta)$ and the corresponding current excitation are expressed as
\begin{equation}
    D_0^{\mathrm{opt}}(\phi, \theta) =  \boldsymbol{a}_0^{\mathrm{T}} \mathbf{R}_0^{-1} \boldsymbol{a}_0^* , \quad j_0^{\mathrm{opt}} =  \mathbf{R}_0^{-1} \boldsymbol{a}_0^*.
\end{equation}
Ideally, if it is assumed that there is a geometric structure that can achieve spatial independence between each antenna element, %i.e., ${||\boldsymbol{r}_n - \boldsymbol{r}_m ||}_2 = \frac{q \pi}{2},q \in \mathbb{Z}^+,\forall n,m=1,2,\ldots,N$
i.e. $\mathbf{R}_0 = \mathbf{I}_N$, where $\mathbf{I}_N \in \mathbb{R}^{N \times N}$ is the identity matrix.
In this case, the maximum directivity $D_0^{\mathrm{opt}}(\phi, \theta) = N$.
%This value is independent of factors such as direction and only relates to the number of antennas. 
Therefore, we take this value as a reference for comparison and leverage the concept of reference directivity
\begin{equation}
    D_0^{\mathrm{rel}}(\phi, \theta) = \frac{1}{N} D_0^{\mathrm{opt}}(\phi, \theta) =  \frac{1}{N}\boldsymbol{a}_0^{\mathrm{T}} \mathbf{R}_0^{-1} \boldsymbol{a}_0^* ,
\end{equation}
which better reflects the impact of factors other than the number of antennas on directivity.
This normalization removes the trivial scaling with $N$ and allows us to focus on the intrinsic directional efficiency of the array.
%In addition, this physical quantity can also be understood as the average directivity contributed by a single antenna to the array.
Unlike conventional directivity, which is an absolute measure referenced to the total radiated power, the modified $D_0^{\mathrm{rel}}(\phi, \theta)$ is a normalized metric that reflects only the directional enhancement introduced by the array.
It differs from normalized radiation intensity, which describes only the pattern shape, and from array gain, which is typically associated with coherent combining performance.

\subsection{Considering MC}
%In an array that is closer to practical conditions, the MC between antenna elements cannot be ignored. 
The effect of MC on the array can be modeled via a coupling matrix $\mathbf{C}\in \mathbb{C}^{N \times N}$ \cite{9838533}
\begin{equation}
    \boldsymbol{j}_c = \mathbf{C} \boldsymbol{j}_0,
\end{equation}
where ${[\mathbf{C}]}_{n,m}$ represents the MC coefficient between the $n$-th antenna and the $m$-th antenna.
The conversion relationship between the MC matrix and the impedance matrix is as follows at the transmitting (TX) and receiving (RX) ends, respectively \cite{1167256, 8058474}
\begin{equation}
    \mathbf{C}_T=\left(1+z_S / z_A\right) \mathbf{Z}\left(\mathbf{Z}+z_S \mathbf{I}_N\right)^{-1} ,
\end{equation}
\begin{equation}
    \mathbf{C}_R=\left(z_A+z_L\right)\left(\mathbf{Z}+z_L \mathbf{I}_N\right)^{-1} ,
\end{equation}
where $\mathbf{Z}, z_A, z_S, z_L$ denote impedance matrix, antenna self-impedance, source impedance, and load impedance, respectively.
The MC values are related to impedance and correspond to the specific type of antenna.
In this letter, we consider electric dipole antennas with length $l$.
The antennas are aligned along the $x$-axis, so the relationship between the antennas is side by side. 
In this case, the mutual impedance $Z_{nm} =  R_{nm} + j X_{nm}$ is \cite{balanis2016antenna}
\begin{equation}
    \begin{aligned}
        R_{nm} & =\frac{\eta}{4 \pi}\left[2 C_i\left(u_0\right)-C_i\left(u_1\right)-C_i\left(u_2\right)\right] , \\
        X_{nm} & =-\frac{\eta}{4 \pi}\left[2 S_i\left(u_0\right)-S_i\left(u_1\right)-S_i\left(u_2\right)\right]  ,\\
        u_0 & =k {||\boldsymbol{r}_n - \boldsymbol{r}_m ||}_2  ,\\
        u_1 & =k\left(\sqrt{{||\boldsymbol{r}_n - \boldsymbol{r}_m ||}_2^2+l^2}+l\right)  ,\\
        u_2 & =k\left(\sqrt{{||\boldsymbol{r}_n - \boldsymbol{r}_m ||}_2^2+l^2}-l\right) ,
    \end{aligned} \label{MC_Calculate}
\end{equation} 
where $\eta$ denotes the intrinsic impedance of the propagation medium and $S_i(x)=\int_0^x \frac{\sin (\tau)}{\tau} \mathrm{d} \tau$ and $C_i(x)=\int_{\infty}^x \frac{\cos (\tau)}{\tau} \mathrm{d} \tau$ are the sine and cosine integrals.

Considering MC, the response of the array is written as
\begin{equation}
    J_c(\phi, \theta) = \boldsymbol{a}_0^{\mathrm{T}} \boldsymbol{j}_c = \boldsymbol{a}_0^{\mathrm{T}} \mathbf{C} \boldsymbol{j}_0  = \boldsymbol{a}_c^{\mathrm{T}} \boldsymbol{j}_0, 
\end{equation}
where $\boldsymbol{a}_c = \mathbf{C}^{\mathrm{T}} \boldsymbol{a}_0$ is effective array response vector.
The spatial correlation matrix considering MC is \cite{Sun22,artemova2021mutual}
\begin{equation}
    \mathbf{R}_c =  \mathbf{C}^{\mathrm{T}} \mathbf{R}_0 \mathbf{C}^*. \label{SpatialCorrelationMatrixMC}
\end{equation}
Under full-space isotropic scattering, \eqref{small_curvature_spacing} and \eqref{SpatialCorrelationMatrix} give $\mathbf R_0(\beta)-\mathbf R_0(0)=\mathcal O(\beta^2)$.
Combining this with \eqref{SpatialCorrelationMatrixMC}, the Hermitian eigenvalue perturbation bound yields
\begin{equation}
 \begin{aligned}
 &|\lambda_i(\mathbf R_c(\beta))-\lambda_i(\mathbf R_c(0))|\\
 &\qquad\leq\|\mathbf R_c(\beta)-\mathbf R_c(0)\|_2
 =\mathcal O(\beta^2),
 \end{aligned}
 \label{curvature_mode_bound}
\end{equation}
where $\lambda_i$ is the $i$th eigenvalue in descending order and $\|\cdot\|_2$ is the matrix spectral norm.
This local bound links curvature-dependent MC to spatial-mode strengths; it does not impose a universal sign on their changes or cover arbitrary angular profiles.

%In order to reflect the difference in maximum directivity, the directivity incorporating MC is defined as
To isolate the impact of MC on the directional response, we define a coupling-aware reference directivity metric by using the uncoupled radiated power as a common reference
\begin{equation}
    D_c\left(\boldsymbol{j}_0, \phi_0, \theta_0 \right)=\frac{4 \pi\left|J_c\left(\phi_0, \theta_0 \right)\right|^2}{\int_0^\pi \int_0^{2\pi}\left|J_0\left(\phi, \theta \right)\right|^2 \sin \theta \mathrm{d} \phi \mathrm{d} \theta},
\end{equation}
from which the simplified expression and reference directivity can be derived as
\begin{equation}
    D_c\left(\boldsymbol{j}_0, \phi, \theta \right)=\frac{\boldsymbol{j}_c^{\mathrm{H}} \boldsymbol{a}_0^* \boldsymbol{a}_0^{\mathrm{T}} \boldsymbol{j}_c}{\boldsymbol{j}_0^{\mathrm{H}} \mathbf{R}_0 \boldsymbol{j}_0} = \frac{\boldsymbol{j}_0^{\mathrm{H}} \mathbf{C}^{\mathrm{H}} \boldsymbol{a}_0^* \boldsymbol{a}_0^{\mathrm{T}} \mathbf{C} \boldsymbol{j}_0}{\boldsymbol{j}_0^{\mathrm{H}} \mathbf{R}_0 \boldsymbol{j}_0} ,
\end{equation}
\begin{equation}
    D_c^{\mathrm{rel}}(\phi, \theta) = \frac{1}{N} D_c^{\mathrm{opt}}(\phi, \theta) =  \frac{1}{N}\boldsymbol{a}_0^{\mathrm{T}} \mathbf{C} \mathbf{R}_0^{-1} \mathbf{C}^{\mathrm{H}} \boldsymbol{a}_0^* .
\end{equation}
This metric is not intended to represent conventional antenna directivity with MC. 
If the denominator were also evaluated using $\mathbf C^H \mathbf R_0 \mathbf C$, the optimized maximum value would be identical to the uncoupled case for nonsingular $\mathbf C$, with MC only changing the optimal excitation. 
Therefore, the uncoupled denominator is retained as a common reference to isolate the MC-induced change in directional response.
For a same-excitation comparison with $N=128$, $L=0.315$ m, $\beta=\pi/8$, $(\phi,\theta)=(\pi/2,\pi/3)$, and $\boldsymbol j_0=\boldsymbol a_0^*/\sqrt N$, the reference metric is 76.744, whereas the conventional coupled directivity obtained using the coupled radiated-power denominator is 63.416; their ratio, 1.2102, equals the coupled-to-reference radiated-power ratio.

For a link, let $\mathbf H_0\in\mathbb C^{N_R\times N_T}$ denote the uncoupled channel and $\mathbf H_c=\mathbf C_R\mathbf H_0\mathbf C_T$ its coupled counterpart \cite{8058474}, with $N_T$ TX and $N_R$ RX elements.
Equal-power Gaussian signaling gives the spectral efficiency \cite{foschini1998limits}
\begin{equation}
 \mathcal R=\log_2\left|\mathbf I_{N_R}+\frac{\rho}{N_T}
 \mathbf R_n^{-1/2}\mathbf H_c\mathbf H_c^{\mathrm H}
 \mathbf R_n^{-1/2}\right|\quad[\mathrm{bit/s/Hz}],
 \label{spectral_efficiency_colored}
\end{equation}
where $\mathbf I_{N_R}$ is the identity, $\rho$ is total model-input signal power relative to reference per-port noise power, and $\mathbf R_n$ is the positive-definite noise covariance in these units.
The Hermitian inverse square root $\mathbf R_n^{-1/2}$ whitens the noise; $\mathbf R_n=\mathbf I_{N_R}$ gives the white-noise baseline.
For no MC, replace $\mathbf H_c$ by $\mathbf H_0$.
Equivalently, spectral efficiency sums $\log_2(1+\rho\mu_i/N_T)$ over the eigenvalues $\mu_i$ of the whitened channel Gram matrix $\mathbf R_n^{-1/2}\mathbf H_c\mathbf H_c^{\mathrm H}\mathbf R_n^{-1/2}$.
Thus, changes in spatial-mode strengths and noise covariance both enter the performance comparison; $\rho$ is not the post-MC receive SNR.

\section{Numerical Results and Analysis}
In this section, the directivity, DoF, and spectral efficiency of 1D HoloCuRAs with different geometries are analyzed both with and without MC.
We set $L=0.315\,\mathrm{m},f=30\,\mathrm{GHz}$, where $f$ is the carrier frequency.
With the fixed aperture length, the inter-element spacings are approximately $0.5\lambda$ for $N=64$ and $0.25\lambda$ for $N=128$, respectively.
Among the parameters related to MC, $\eta = 120\pi \,\mathrm{\Omega}$ for free-space propagation, $z_A = 73 + j42.5 \, \mathrm{\Omega}$ for $l = \frac{\lambda}{2}$, and $z_S = z_L = z_A^*$ for conjugate matching.
In this work, we restrict $\beta \leq \pi / 8$ to focus on small geometric deformations relevant to practical curved deployments.
%In this work, we focus on a HoloCuRA subject to only small geometric deformations due to mechanical stress or temperature variations in many practical applications, such that the angular deviation $\beta$ is restricted to be no larger than $\frac{\pi}{8} \approx 0.39$.
%This bound follows a commonly adopted engineering rule-of-thumb, which is widely used in array and antenna theory to ensure the validity of small-angle and planar-wave approximations.

\subsection{Directivity}
%When the excitation is uniform current and MC is not considered, the beam pattern of the flexible ULAs is shown in Fig.~\ref{FigBeamPattern}.
%\begin{figure}[htbp]
%	\centering
%	\includegraphics[scale=0.80]{figures/BeamPattern.png}
%	\caption{Beam pattern in the $yz$ plane of the flexible ULAs with different curvatures, where $N=64$.}  % 主图标题
%	\label{FigBeamPattern} % 主图标签
%\end{figure}
%A clear trend is observed: as the arch curvature increases, the main-lobe peak decreases monotonically, while sidelobe peaks rise and broaden; once a curvature threshold is exceeded, the main lobe splits.
%With further curvature, the beampattern exhibits multiple dominant peaks of comparable amplitude, indicating a more dispersed angular energy distribution. 

\begin{figure*}[htbp]
    \centering
    % 第一行
    \subfloat[]{%
        \includegraphics[width=0.2\linewidth]{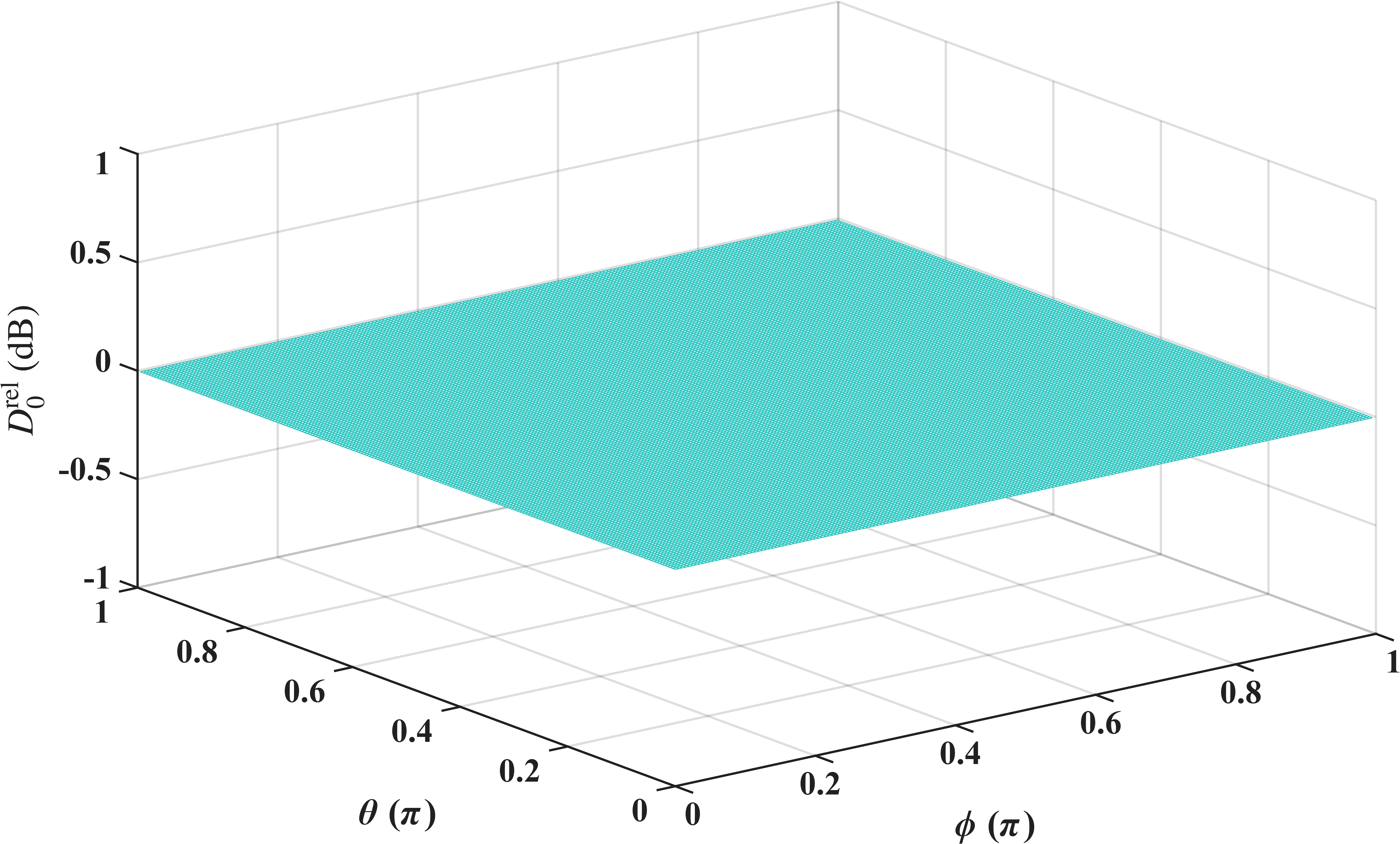}%
        \label{figRelDirecBeta0.00Uncoupled64}}
    \hfill
    \subfloat[]{%
        \includegraphics[width=0.2\linewidth]{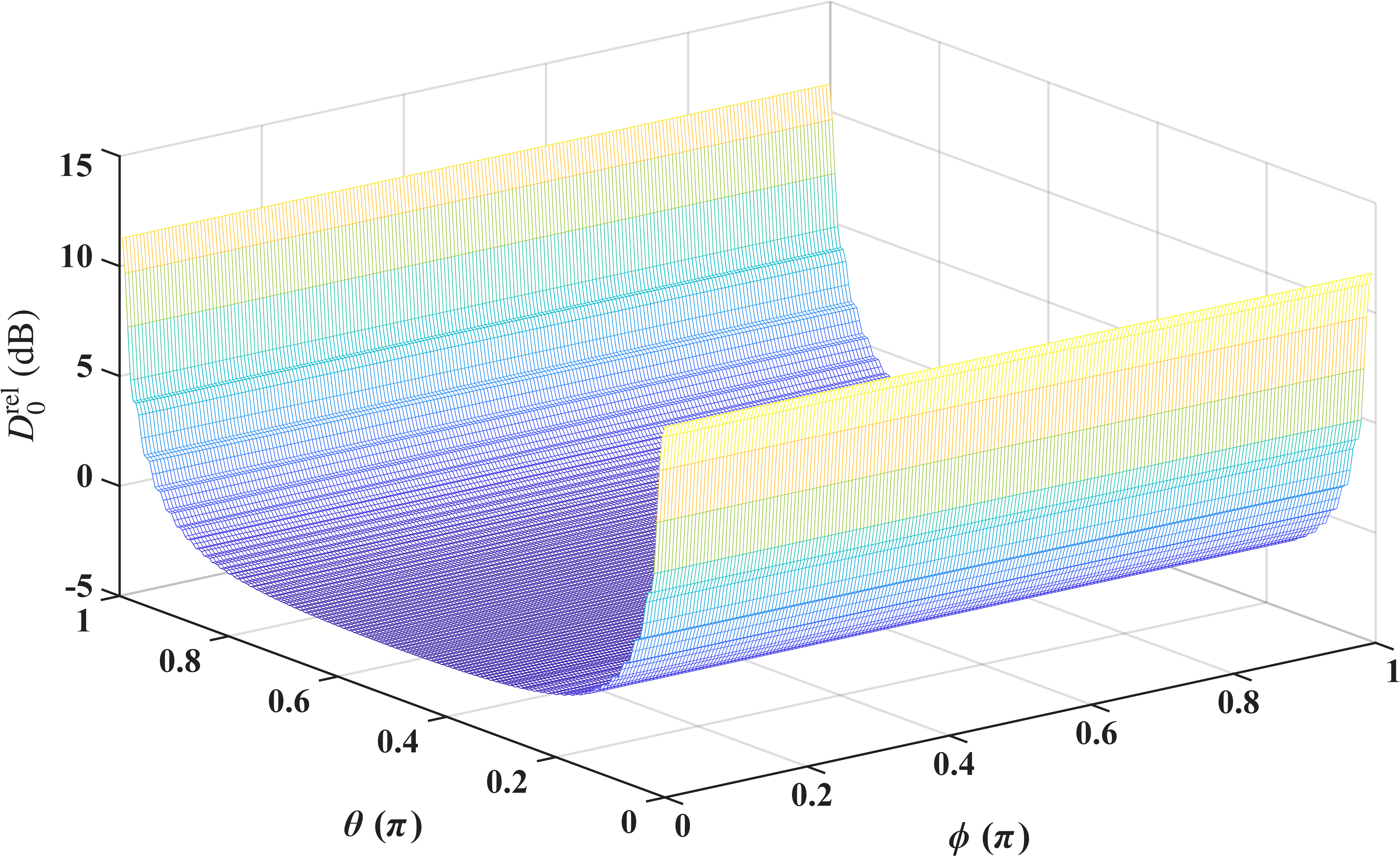}%
        \label{figRelDirecBeta0.00Uncoupled}}
    \hfill
    \subfloat[]{%
        \includegraphics[width=0.2\linewidth]{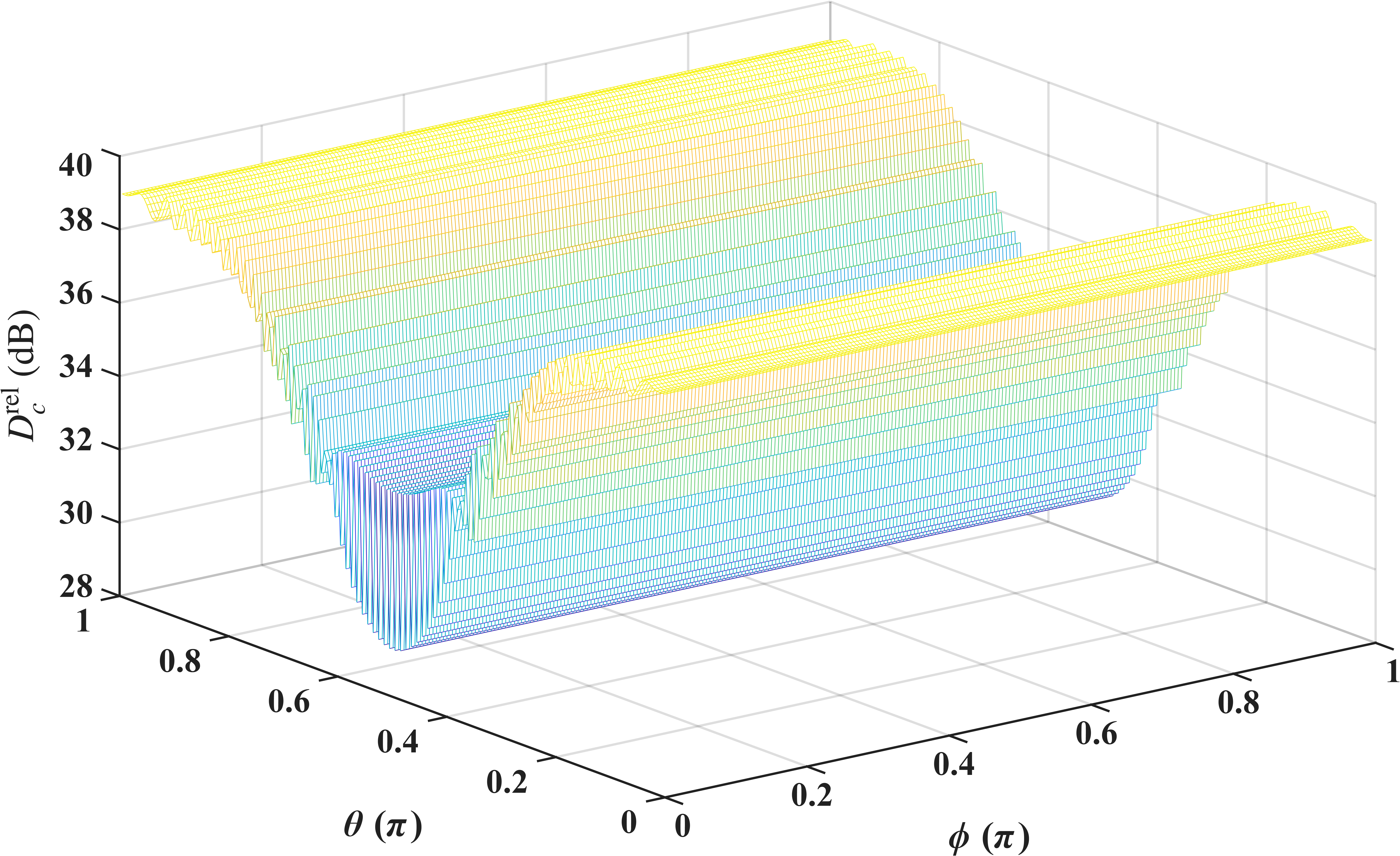}%
        \label{figRelDirecBeta0.00Coupled}}
    \hfill
    \subfloat[]{%
        \includegraphics[width=0.2\linewidth]{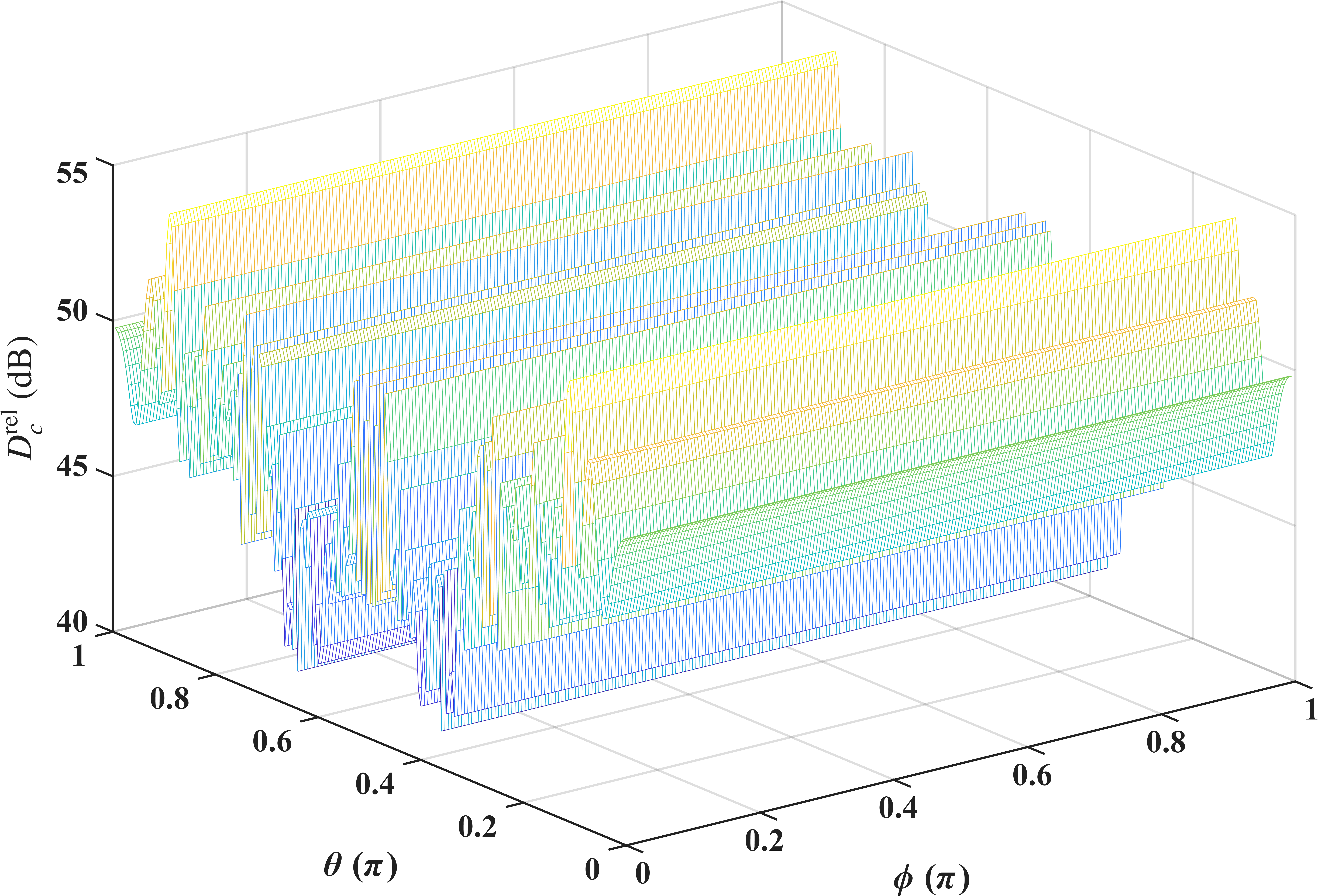}%
        \label{figRelDirecBeta0.00Air}}
    \hfill
    \subfloat[]{%
        \includegraphics[width=0.2\linewidth]{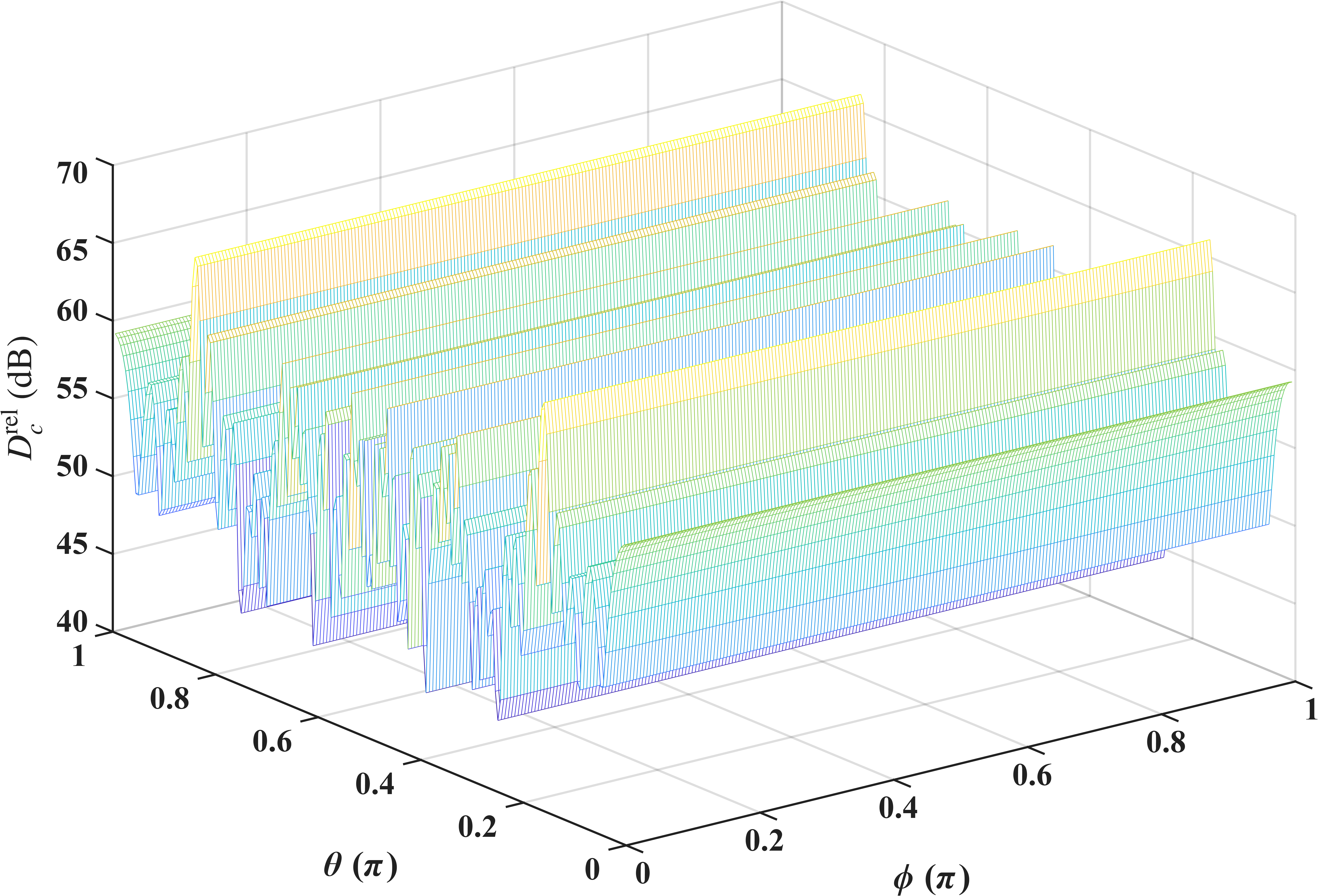}%
        \label{figRelDirecBeta0.00Back}}
    \vspace{0.005em}
    \subfloat[]{%
        \includegraphics[width=0.2\linewidth]{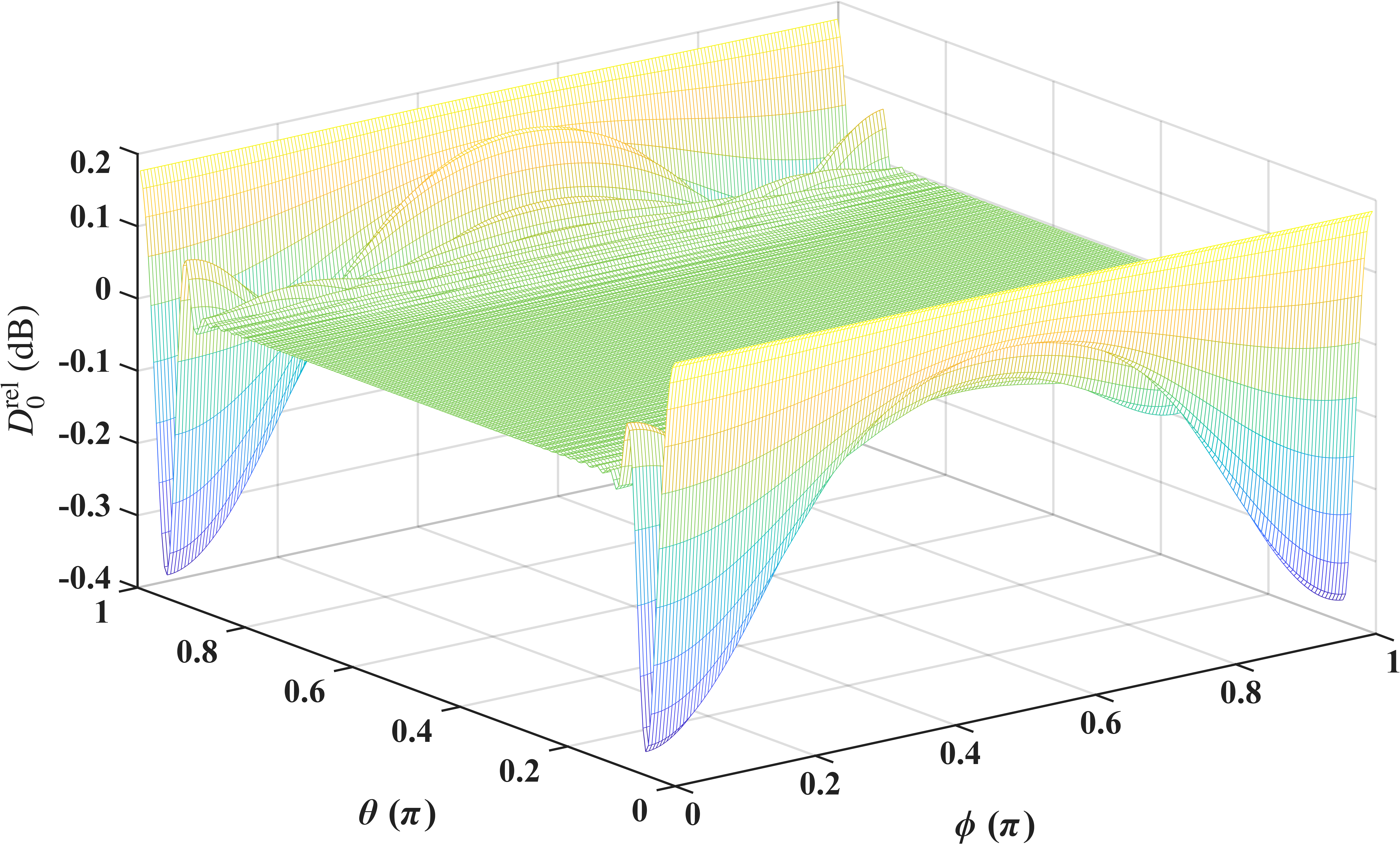}%
        \label{figRelDirecBeta0.39Uncoupled64}}
    \hfill
    \subfloat[]{%
        \includegraphics[width=0.2\linewidth]{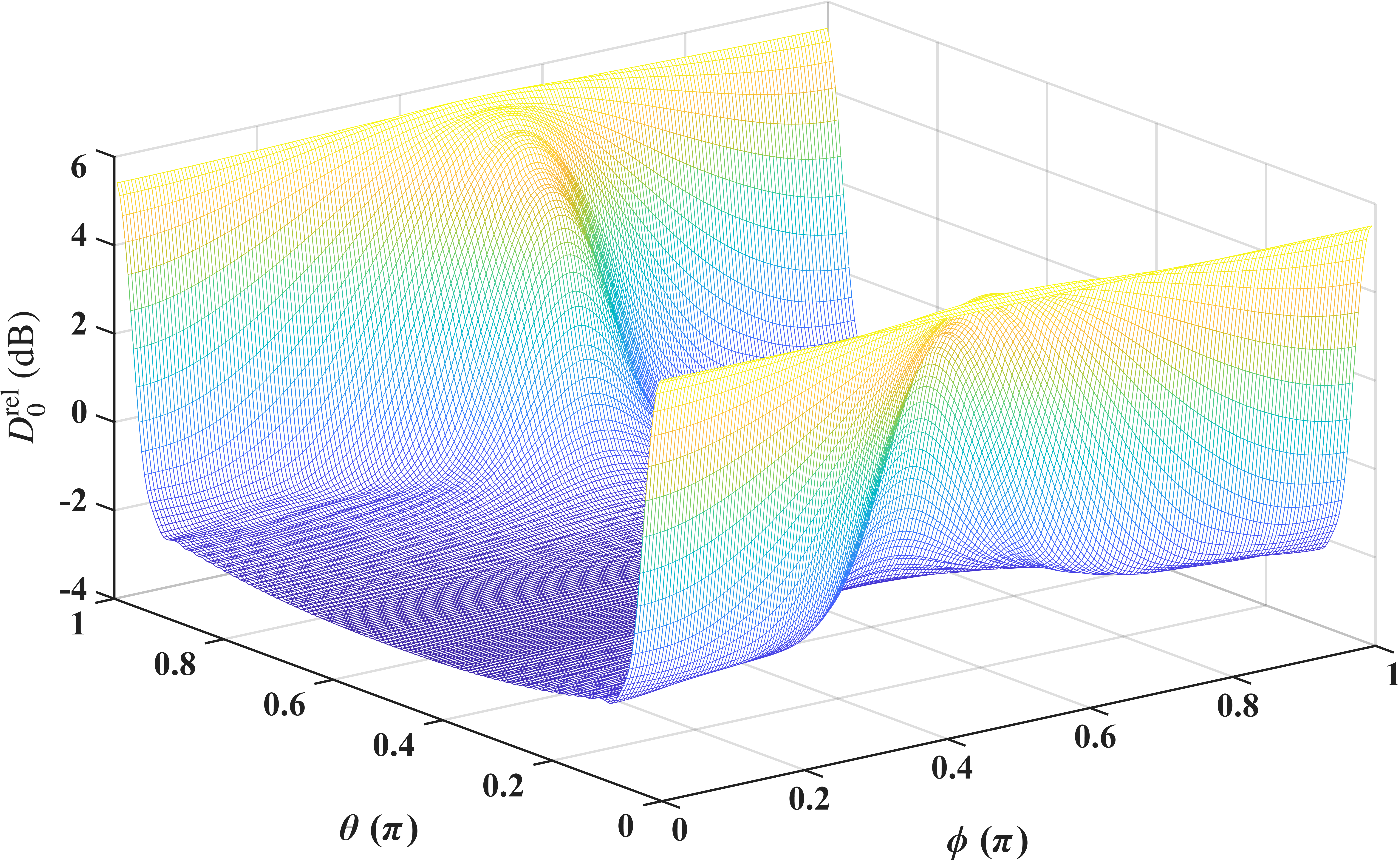}%
        \label{figRelDirecBeta0.39Uncoupled}}
    \hfill
    \subfloat[]{%
        \includegraphics[width=0.2\linewidth]{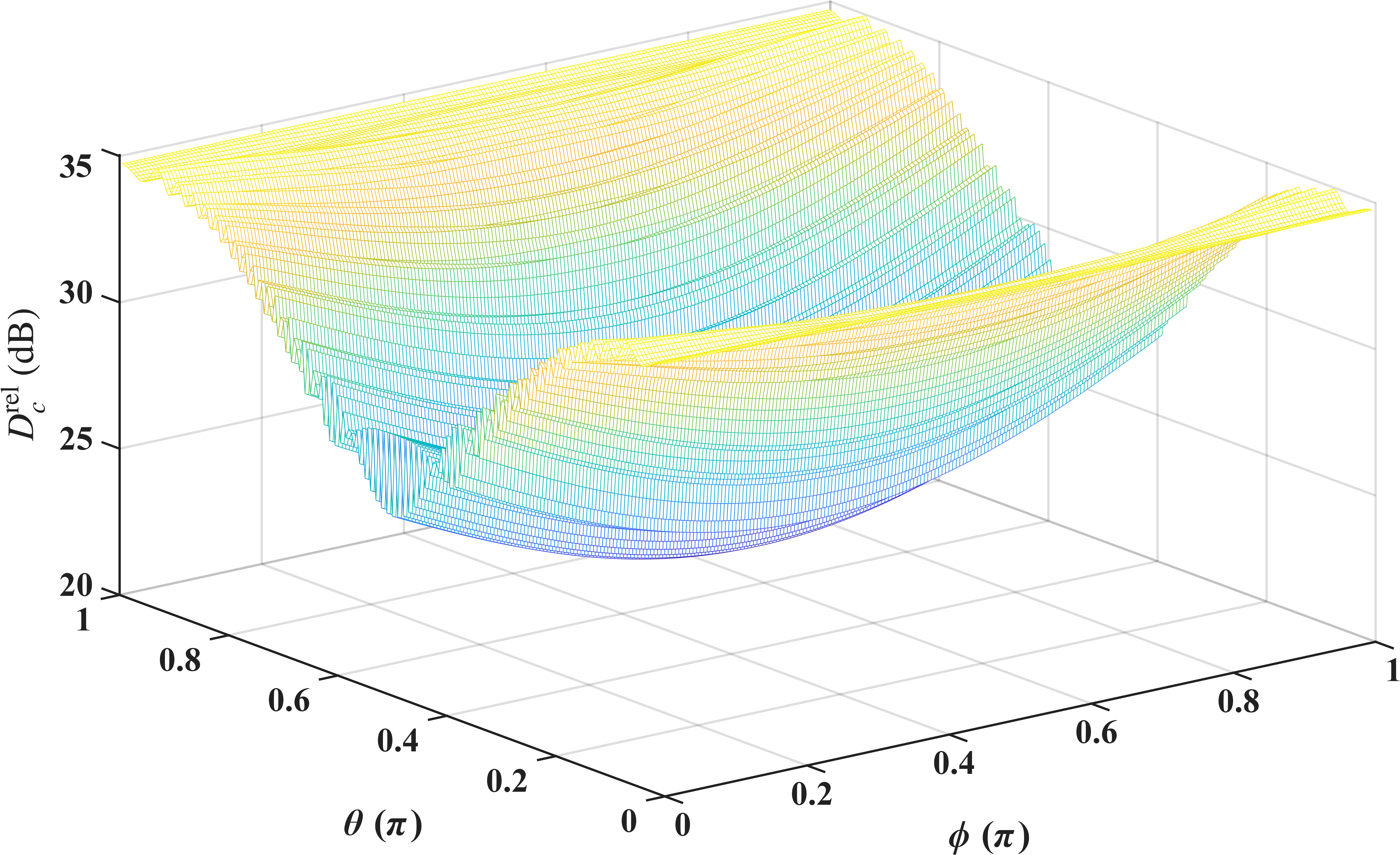}%
        \label{figRelDirecBeta0.39Coupled}}
    \hfill
    \subfloat[]{%
        \includegraphics[width=0.2\linewidth]{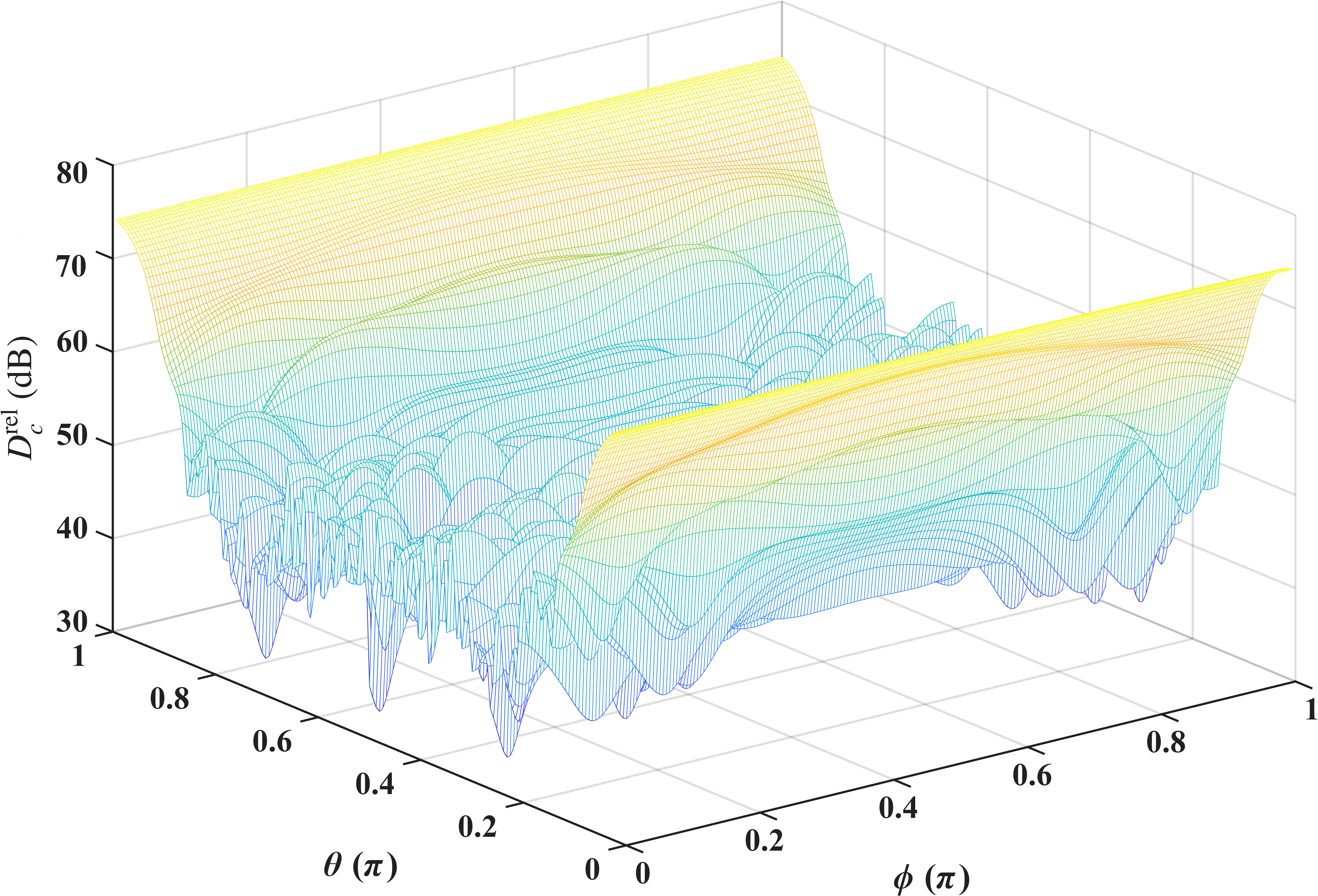}%
        \label{figRelDirecBeta0.39Air}}
    \hfill
    \subfloat[]{%
        \includegraphics[width=0.2\linewidth]{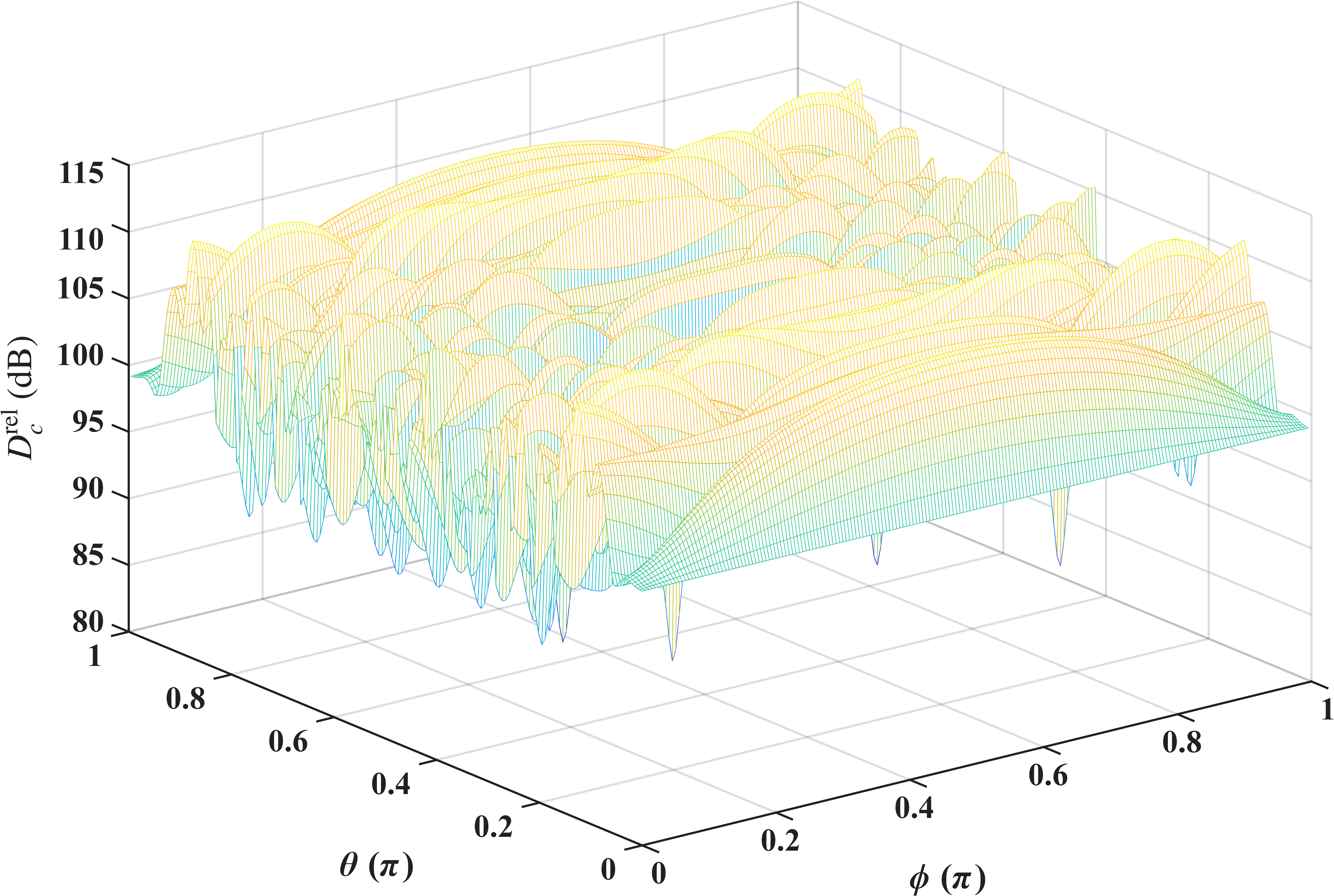}%
        \label{figRelDirecBeta0.39Back}}
	\captionsetup{font=footnotesize}
    \caption{Reference directivity of HoloCuRAs when ignoring and considering MC. Subfigures (a) and (f) show the $N=64$ reference cases, while the remaining subfigures correspond to $N=128$. (a) $\beta=0.00$, ignoring MC. (b) $\beta=0.00$, ignoring MC. (c) $\beta=0.00$, MC calculated using \eqref{MC_Calculate}. (d) $\beta=0.00$, dipoles in a vacuum. (e) $\beta=0.00$, printed dipoles on the substrate. (f) $\beta=0.39$, ignoring MC. (g) $\beta=0.39$, ignoring MC. (h) $\beta=0.39$, MC calculated using \eqref{MC_Calculate}. (i) $\beta=0.39$, dipoles in a vacuum. (j) $\beta=0.39$, printed dipoles on the substrate.}
    \label{FigRelativeDirectivity}
\end{figure*}

The reference directivity patterns of HoloCuRAs under different curvature conditions are shown in Fig.~\ref{FigRelativeDirectivity}.
The spatial correlation matrix is derived under full-space isotropic scattering, while Fig.~\ref{FigRelativeDirectivity} focuses on the forward angular region $\phi \in [0,\pi], \theta \in [0,\pi]$, consistent with the $+y$-oriented HoloCuRA deployment.
The $N=64$ cases (Fig.~\ref{FigRelativeDirectivity}(a)(f)) are provided as reference examples, while the remaining results are obtained with $N=128$.
The coupling matrix can be obtained either from the analytical expression in \eqref{MC_Calculate} or from full-wave electromagnetic simulations.
In this paper, the full-wave simulation results are generated under two different physical modeling scenarios.
The first scenario considers an array of ideal cylindrical dipoles in free space, as shown in Fig.~\ref{FigRelativeDirectivity}(d)(i), which is consistent with the assumptions adopted in the theoretical derivation.
The second scenario considers printed dipoles mounted above a perfect electric conductor (PEC) reflector using a Rogers RT/Duroid 5880 substrate \cite{7926309,4907127}, as illustrated in Fig.~\ref{FigRelativeDirectivity}(e)(j).
Although this configuration introduces practical effects such as dielectric loading and loss, it provides a more realistic representation.
In addition, under curved configurations, the effects of element blockage and shadowing are inherently taken into account.

As the array curvature increases, the reference directivity exhibits more pronounced angular redistribution and spatial non-uniformity. 
When MC is considered, the pattern changes significantly, especially along the zenith-angle dimension, indicating that MC reshapes the angular response of densely deployed HoloCuRAs. 
The cylindrical-dipole full-wave results follow the analytical trends, with magnitude differences mainly caused by finite dipole thickness and numerical electromagnetic effects.

Compared with cylindrical dipoles in free space, the reflector-backed printed dipoles show more complex pattern fluctuations due to dielectric loading, reflector effects, and possible blockage or shadowing. 
Nevertheless, both full-wave cases exhibit clear curvature-induced angular redistribution, confirming that array curvature plays a key role in shaping the radiation behavior of HoloCuRAs.

\subsection{DoF and Spectral Efficiency}
\begin{figure}[t]
    \centering
    \subfloat[]{%
        \includegraphics[width=0.50\linewidth]{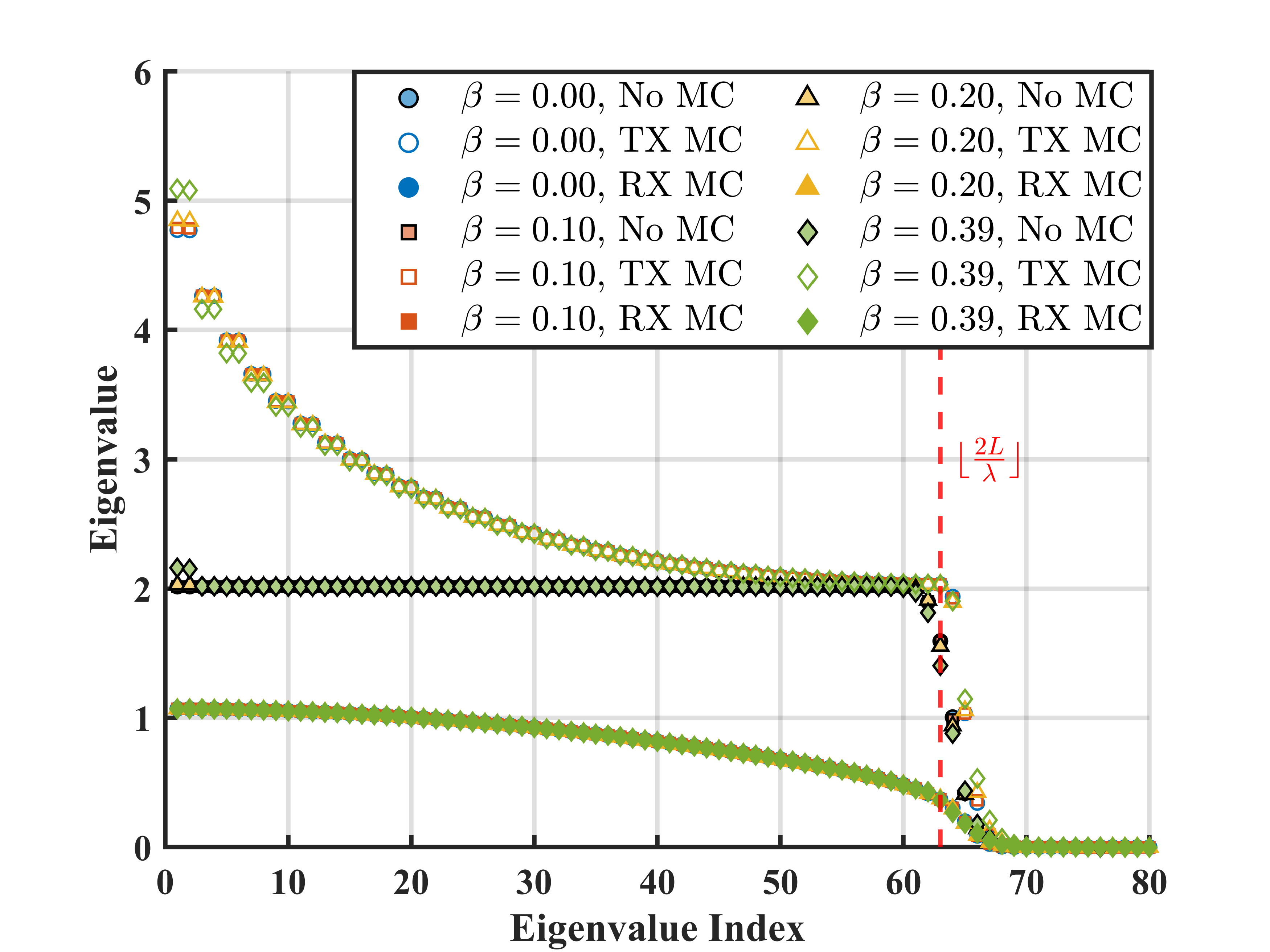}%
        \label{figIsotropicDoF}}
    \hfill
    \subfloat[]{%
        \includegraphics[width=0.50\linewidth]{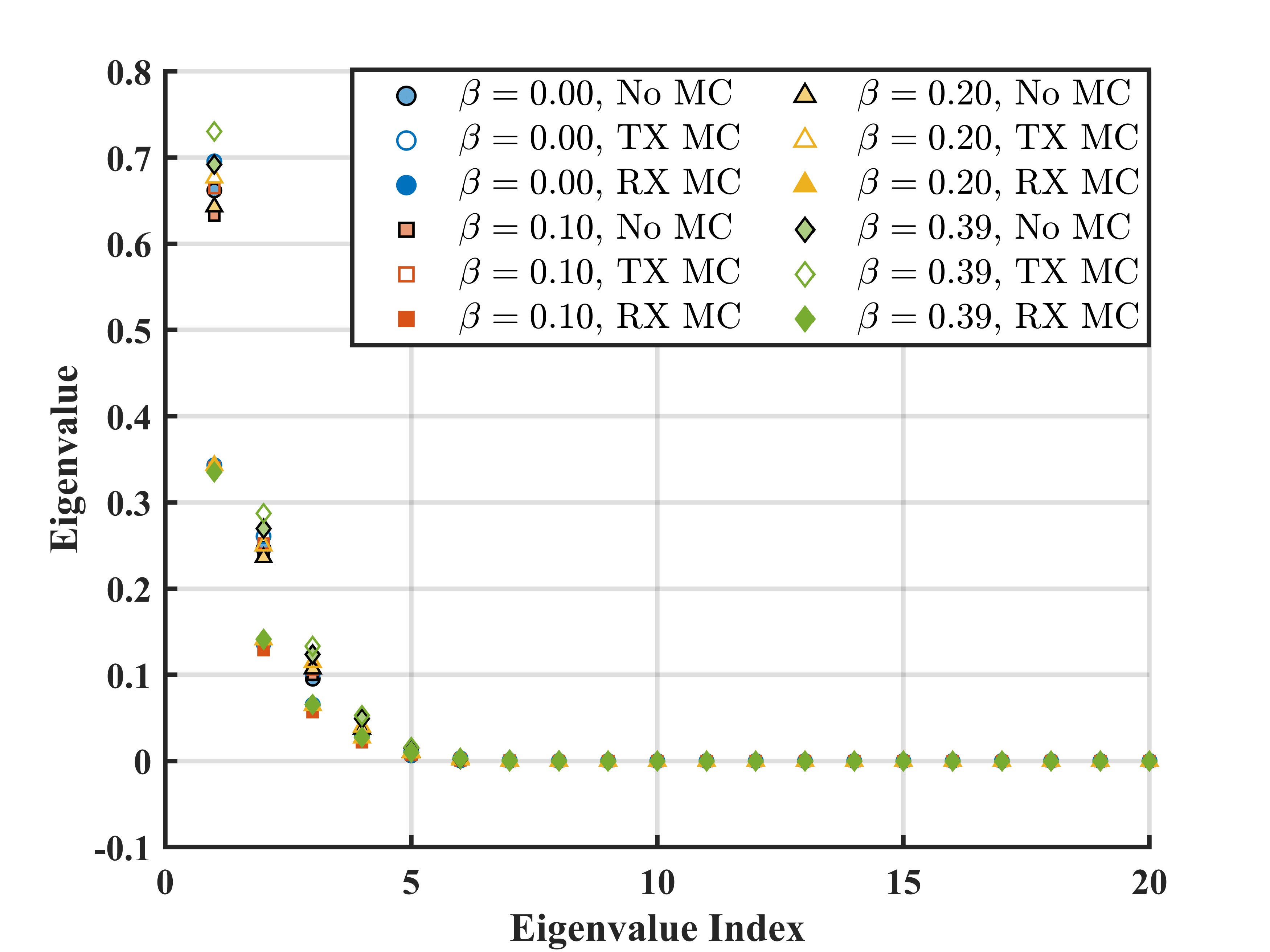}%
        \label{figDoFCDLB}}
    \captionsetup{font=footnotesize}
    \caption{Eigenvalues of HoloCuRAs ignoring and considering MC with different curvatures, where $N=128$. (a) Isotropic scattering. (b) CDL-B.}
    \label{FigDoF}
\end{figure}

Under isotropic scattering, the dominant eigenvalues of the array spatial correlation matrix are shown in Fig.~\ref{FigDoF}(a), while Fig.~\ref{FigDoF}(b) gives the corresponding CDL-B result.

Within $\beta\leq\pi/8$, curvature has limited influence on the dominant eigenvalue count, whereas MC changes the mode strengths.
TX coupling generally raises the dominant eigenvalues under the adopted pre-MC channel normalization, while RX coupling reduces them.
Eigenvalues beyond the reference index $\lfloor2L/\lambda\rfloor$ rapidly approach similar small values, so this index remains a useful indicator of the available spatial modes in the considered regime.

\begin{figure}[t]
    \centering
    \includegraphics[width=\columnwidth]{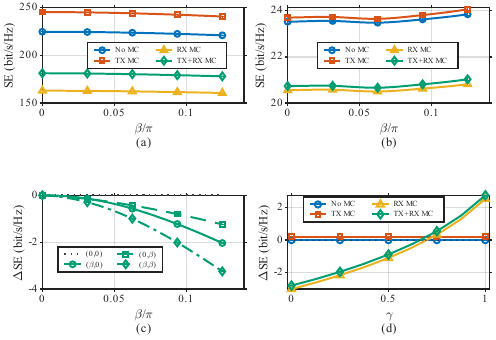}
    \captionsetup{font=footnotesize}
    \caption{Spectral efficiency for $N_T=N_R=128$ and $\rho=10$ dB. (a) Hemispherical isotropic and (b) CDL-B profiles with symmetric curvature. (c) Isotropic TX+RX MC: changes relative to coupled planar arrays. (d) CDL-B at $\beta_T=\beta_R=\pi/8$: changes relative to no MC versus noise fraction $\gamma$. Selected markers show 95\% confidence intervals from 1000 paired realizations.}
    \label{FigSE}
\end{figure}

Figure~\ref{FigSE} uses $N_T=N_R=128$, aperture lengths $L_T=L_R=0.315$ m, $\rho=10$ dB, and independent TX/RX curvature angles $\beta_T,\beta_R\in[0,\pi/8]$.
Each realization satisfies $\|\mathbf H_0\|_F=\sqrt{N_TN_R}$ before MC, without renormalizing $\mathbf H_c$; the comparison fixes model-input rather than radiated power.
Hemispherical scattering represents two opposing reflector-backed arrays, whereas CDL-B uses visible cluster-center angles and powers in a model.
All comparisons reuse random draws across configurations and use 1000 realizations; paired 95\% confidence-interval half-widths equal 1.96 standard errors.
Panels (a)--(c) assume independent white downstream noise.

Symmetric bending in Fig.~\ref{FigSE}(a) reduces spectral efficiency under isotropic scattering, whereas it increases spectral efficiency under CDL-B in (b).
From zero curvature to $\pi/8$, the TX+RX-MC changes are $-3.246$ and $+0.293$ bit/s/Hz, respectively, compared with $-3.732$ and $+0.319$ bit/s/Hz without MC.
TX MC raises spectral efficiency and RX MC lowers it under the adopted normalization and white-noise baseline.
The contrasting trends show the dependence on the angular power distribution.
The isotropic coupled-minus-uncoupled bending contrast is $0.486\pm0.006$ bit/s/Hz.
A fixed-planar-MC control attributes only $-0.01369\pm0.00090$ bit/s/Hz to updating the coupling matrices, indicating that the main effect is the network-dependent response to geometry.

In Fig.~\ref{FigSE}(c), all four configurations include TX+RX MC, and $\Delta\mathrm{SE}$ denotes the change from the coupled planar case.
At $\beta=\pi/8$, the changes for $(\beta,0)$, $(0,\beta)$, and $(\beta,\beta)$ are $-2.036\pm0.014$, $-1.241\pm0.012$, and $-3.246\pm0.019$ bit/s/Hz, respectively, where $\pm$ gives the 95\% interval half-width.
RX-only bending has a smaller loss by $0.795\pm0.019$ bit/s/Hz; the unplotted no-MC TX-minus-RX control is $-0.002\pm0.021$ bit/s/Hz.
Thus, the distinct source/load transformations make the bending response depend on the link end under this baseline.
The zero-valued $(0,0)$ reference and one-sided curves isolate bending while retaining MC at both ends.

Coupling also shapes receiver noise \cite{dong2009noise}.
Let $\sigma_e^2$ and $\sigma_i^2$ denote per-port noise powers before and after the receive network.
Independent white sources then give covariance $\sigma_e^2\mathbf C_R\mathbf C_R^{\mathrm H}+\sigma_i^2\mathbf I_{N_R}$ before output-power normalization.
Fixing average output-noise power gives the Fig.~\ref{FigSE}(d) model
\begin{equation}
 \mathbf R_n(\gamma)=(1-\gamma)\mathbf I_{N_R}
 +\gamma\frac{N_R\mathbf C_R\mathbf C_R^{\mathrm H}}
 {\operatorname{tr}(\mathbf C_R\mathbf C_R^{\mathrm H})},
 \label{noise_sensitivity}
\end{equation}
where $\operatorname{tr}(\cdot)$ denotes trace and $\gamma\in[0,1]$ is the pre-network source's fraction of average output-noise power; $\mathbf C_R=\mathbf I_{N_R}$ without RX MC.
As $\gamma$ increases from 0 to 1, the RX-MC and TX+RX-MC differences relative to no MC change from $-3.022$ and $-2.809$ to $+2.555$ and $+2.770$ bit/s/Hz; TX MC remains at approximately $+0.215$ bit/s/Hz.
This covariance-shape sensitivity is not a complete antenna/LNA thermal-noise model: the positive endpoint does not imply a universal RX-coupling gain.

\section{Conclusion}
This letter characterized the spatial correlation, coupling-aware reference directivity, and spectral efficiency of HoloCuRAs with MC.
The small-curvature relation links geometry to coupling and spatial-mode perturbations under isotropic scattering.
Within the tested configurations, MC produces unequal spectral-efficiency responses to TX and RX array curvature under the white-noise baseline, while the no-MC difference is consistent with zero under isotropic scattering.
The contrasting isotropic and CDL-B trends, together with the colored-noise results, show that array curvature and RX coupling cannot be assigned a universal benefit or penalty.
Curved-array evaluation should therefore account jointly for geometry, coupling, angular power distribution, and receiver-noise covariance.

\bibliographystyle{IEEEtran}  % 设置IEEE引用样式
\bibliography{references}

\end{document}